\documentclass{aa}
\usepackage{graphics}
\usepackage{supertabular}

\begin {document}
\title{The NASA Astrophysics Data System: Data Holdings}

\thesaurus{04(04.01.1)}
\author{C. Grant\and A. Accomazzi\and G. Eichhorn\and M. J. Kurtz
\and S. S. Murray}
\institute{Harvard-Smithsonian Center for Astrophysics, Cambridge, MA 02138}

\offprints{C. Grant}
\mail{C. Grant}

\date{Received / Accepted}

\titlerunning{}
\authorrunning{C. Grant et al.}

\maketitle

\sloppy

\begin {abstract}
  Since its inception in 1993, the ADS Abstract Service has become an
indispensable research tool for astronomers and astrophysicists worldwide.
In those seven years, much effort has been directed toward improving
both the quantity and the quality of references in the database.  From the 
original database of approximately 160,000 astronomy abstracts, our
dataset has grown almost tenfold to approximately 1.5 million references
covering astronomy, astrophysics, planetary sciences, physics, optics, and
engineering.  We collect and standardize data from 
approximately 200 journals and present the resulting information in a
uniform, coherent manner.  With the cooperation of journal publishers 
worldwide, we have been able to place scans of full journal articles on-line
back to the first volumes of many astronomical journals, and we are able to 
link to current version of articles, abstracts, and datasets for 
essentially all of the current astronomy literature.   The
trend toward electronic publishing in the field, the use of electronic
submission of abstracts for journal articles and conference proceedings, 
and the increasingly prominent use of the World Wide Web to disseminate 
information have enabled the ADS to build a database unparalleled in other 
disciplines.

The ADS can be accessed at http://adswww.harvard.edu 
\end{abstract}

\section {\label {intro} Introduction}

     Astronomers today are more prolific than ever before.  Studies in
publication trends in astronomy (\cite{1994PASP..106.1015A}, 
\cite{1995ApJ...455..407A}, \cite{1997PASP..109.1278S})
have hypothesized that the current explosion in published papers in astronomy 
is due to a combination of factors:  growth in professional society
membership, an increase in papers by multiple authors, the launching of
new spacecrafts, and increased competition for jobs and PIs in the field
(since candidate evaluation is partially based on publication history).
As the number of papers in the field grows, so does the need for
tools which astronomers can use to locate that fraction of papers which
pertain to their specific interests.

     The ADS Abstract Service is one of several bibliographic services
which provide this function for astronomy, but due to the broad scope of our coverage and 
the simplicity of access to our data, astronomers now rely extensively on the
ADS, and other bibliographic services not only link to us, but some have 
built their bibliographic search capabilities on top of the ADS system.
The International Society for Optical Engineering (SPIE) and the
NASA Technical Report Service (NTRS) are two such services.

     The evolution of the Astrophysics Data System (ADS) has been largely 
data-driven.  Our search tools and indexing routines have been modified
to maximize speed and efficiency based on the content of our dataset.
As new types of data (such as electronic versions of articles) became
available, the Abstract Service quickly incorporated
that new feature.  The organization and standardization of the database
content is the very core upon which the Abstract Service has been built.

     This paper contains a description of the ADS Abstract Service from a 
``data" point of view, specifically descriptions of our holdings and of the
processes by which we ingest new data into the system.  Details are provided 
on the organization of the 
databases (section \ref {databases}), the description of the data in the 
databases (section \ref {data}), the creation of bibliographic records
(section \ref {creating}), the procedures for updating the database
(section \ref {updating}), and on the scanned articles in the Astronomy database 
(section \ref {articles}).  We discuss the interaction between the ADS and
the journal publishers (section \ref {journals}) and analyze some of the numbers 
corresponding to the datasets (section \ref {summary}).
In conjunction with
three other ADS papers in this volume, this paper is intended to offer details on
the entire Abstract Service with the hopes that astronomers will have a 
better understanding of the reference data upon which they rely for their 
research.  In addition, we hope that researchers in other disciplines may
be able to benefit from some of the details described herein.

     As is often the case with descriptions of active Internet resources,
what follows is a description of the present situation with the ADS Abstract
Service.  New features are always being added, some of which necessitate
changes in our current procedures.  Furthermore, with the growth of
electronic publishing, some of our core ideas about bibliographic tools
and requirements must be reconsidered in order to be able to take full
advantage of new publishing technologies for a new millennium.

\section {\label {databases} The Databases}

     The ADS Abstract Service was originally conceived of in the mid 1980's
as a way to provide on-line access to bibliographies of astronomers which 
were previously available only through expensive librarian search services 
or through the A\&A Abstracts series (\cite{1979BICDS..17....2S}, 
\cite{1982adra.proc..159S}, \cite{1989lisa.conf...77S}), published by 
the Astronomisches
Rechen-Institut in Heidelberg.  While the ideas behind the Abstract
Service search engine were being developed (see \cite{kurtz}, hereafter OVERVIEW), 
concurrent efforts were underway to acquire a reliable data source on which 
to build the server.   In order to best develop the logistics of the 
search engine it was necessary to have access to real literature data from
the past and present, and to set up a mechanism for acquiring data in the
future.  

     An electronic publishing meeting in the spring of 1991 brought together
a number of organizations whose ultimate cooperation would be necessary to
make the system a reality (see OVERVIEW for details).  NASA's Scientific and 
Technical Information Program (STI) offered to provide abstracts to the
ADS.  STI's abstracts were a rewritten version of the original abstracts, 
categorized and keyworded by professional editors.  They not only abstracted 
the astronomical literature, but many other scientific disciplines as well.   
With STI agreeable to providing the past and present literature, and the 
journals committed to providing the future literature, the data behind the 
system fell into place.  The termination of the journal abstracting by the 
STI project several years later was unfortunate, but did not cause the 
collapse of the ADS Abstract Service because of the commitment of the 
journal publishers to distribute their information freely.

     The STI abstracting approximately covered the period from 1975 to 1995.
With the STI data alone, we estimated the completeness of the Astronomy database 
to be better than 90\% for the core astronomical journals.  Fortunately, with the additional data supplied by the 
journals, by SIMBAD (Set of Identifications, Measurements, and Bibliographies 
for Astronomical Data, \cite{1988alds.proc..323E}) at 
the CDS (Centre de Donn\'ees Astronomiques de Strasbourg), and by performing 
Optical Character Recognition (OCR) on the scanned table of contents (see 
section \ref{articles} below), we are now closer to 
99\% complete for that period.  In the period since then we are 100\% complete
for those 
journals which provide us with data, and significantly less complete for those 
which do not (e.g. many observatory publications and non-U.S. journals).  
The data prior to 1975 are also 
significantly incomplete, although we are currently working to improve the 
completeness of the early data, primarily through scanning the table 
of contents for journal volumes as they are placed on-line.  We are 100\%
complete for any journal volume which we have scanned and put on-line,
since we verify that we have all bibliographic entries during the procedure of 
putting scans on-line.

     Since the STI data were divided into categories, it was easy to create 
additional databases with non-astronomical data which were still of interest
to astronomers.  The creation of an Instrumentation database has enabled us to
provide a database for literature related to astronomical instrumentation, of 
particular interest to those scientists building astronomical telescopes and
satellite instruments.  We were fortunate to get the cooperation of the 
SPIE very quickly after 
releasing the Instrumentation database.  SPIE has become our major source of
abstracts for the Instrumentation database now that STI no longer supplies us 
with data.  

     Our Physics and Geophysics database, the third database to go on-line,
is intended for scientists working in physics-related fields.  
We add authors and titles from all of the physics journals of the American 
Institute of Physics (AIP), the Institute of Physics (IOP), and the 
American Physical Society (APS), as well as many physics journals 
from publishers such as Elsevier and Academic Press (AP (AP)).  

     The fourth database in the system, the Preprint database, contains a 
subset of the Los Alamos National Laboratory's (LANL) Preprint Archive
(\cite{lanl}).
Our database includes the LANL astro-ph preprints which are retrieved from
LANL and indexed nightly through an automated procedure.  That dataset 
includes preprints from astronomical journals submitted directly by 
authors.

\section {\label {data} Description of the Data}

     The original set of data from STI contained several basic fields of
data (author, title, keywords, and abstracts) to be indexed and made available 
for searching.  All records were keyed on STI's accession number, 
a nine-digit code consisting of a letter prefix (A or N) followed 
by a two-digit publication year, followed by 
a five-letter identifier (e.g. A95-12345).  Data were stored in files
named by accession number.  

With the inclusion of data from other
sources, primarily the journal publishers and SIMBAD, we extended
STI's concept of the accession number to handle other abstracts as well.
Since the ADS may receive the same abstract from multiple sources, we originally
adopted a system of using a different prefix letter with the remainder of
the accession number being the same to describe abstracts received from
different sources.  Thus, the same abstract for the above accession number
from STI would be listed as J95-12345 from the journal publisher and S95-12345 
from SIMBAD.  This allowed the indexing routines to consider only one instance 
of the record when indexing.  Recently, limitations in the format of accession
numbers and the desire to index data from multiple sources (rather than just
STI's version) have prompted us to move to a data storage system based 
entirely on the bibliographic code.

\subsection{\label {bibcodes} Bibliographic Codes}

     The concept of a unique bibliographic code used to identify an article
was originally conceived of by SIMBAD and NED (NASA's Extragalactic Database,
\cite{1988alds.proc..335H}).  The original specification is detailed in
\cite{1995VA.....39R.272S}.  In the years since, the ADS has 
adopted and expanded their definition to be able to describe references 
outside of the scope of those projects.

     The bibliographic code is a 19-character string comprised of several
fields which usually enables a user to identify the full reference from
that string.  It is defined as follows:

\centerline{\bf YYYYJJJJJVVVVMPPPPA}

\noindent
where the fields are defined in Table~\ref{table1}.

\begin{table*}
\caption[]{Bibliographic Code Definition  (e.g. 1996A\&AS..115....1S)}
\label{table1}
\begin{tabular*}{7.0in}{lll}

\hline
\noalign{\smallskip}

Field & Definition & Example

\\
\noalign{\smallskip}
\hline
\noalign{\smallskip}
YYYY  & Publication Year & 1997 \\
JJJJJ & Journal Abbreviation & ApJ, A\&A, MNRAS, etc.   \\
VVVV  & Volume Number & 480 \\
M     & Qualifier for Publication & L (for Letter), P (for Pink Page) \\
& &  Q, R, S, etc. for unduplicating   \\
& &  a, b, c, etc. for issue number   \\
PPPP  & Page Number & 129 \\
A     & First Letter of the First Author's Surname & N \\

\\
\noalign{\smallskip}
\hline
\end{tabular*}
\end{table*}

The journal field is left-justified and the volume and page fields are 
right-justified.  Blank spaces and leading zeroes are replaced by periods.
For articles with page numbers greater than 9999, the M field contains
the first digit of the page number.  The A field contains a colon (``:")
if there is no author listed.

     Creating bibliographic codes for the astronomical journals is 
uncontroversial.  Each journal typically has a commonly-used abbreviation,
and the volume and page are easily assigned (e.g. 1999PASP..111..438F).
Each volume tends to have individual page numbering,
and in those cases where more than one article appears on a page (such as
errata), a ``Q",``R",``S", etc. is used as the qualifier for publication to 
make bibliographic codes unique.  When page numbering is not continuous across issue
numbers (such as Sky \& Telescope), the issue number is represented by
a lower case letter as the qualifier for publication (e.g. ``a" for issue 1).
This is because there may be multiple articles in a volume starting on the same
page number.

     Creating bibliographic codes for the ``grey" literature such as
conference proceedings and technical reports is a more difficult task.
The expansion into these additional types of data included in the ADS 
required us to modify the original prototype bibliographic code definition in 
order to present identifiers which are easily recognizable to the user.  The 
prototype definition of the bibliographic code suggested using a single letter 
in the second place of the volume field to identify non-standard references
(catalogs, PhD theses, reports, preprints, etc.) and using the third and
fourth place of that field to unduplicate and report volume numbers (e.g.
1981CRJS..R.3...14W).  Since we felt that this created codes unidentifiable to 
the typical user and since NED and SIMBAD did not feel that users needed to be 
able to identify books directly from their bibliographic codes, the ADS adopted 
different rules for creating codes to identify the grey literature.

     It is straightforward to create bibliographic codes for conference proceedings which are part of 
a series.  For example, the IAU Symposia Series (IAUS) 
contains volume numbers and therefore fits the journal model for bibliographic 
codes.  Other conference proceedings, books, colloquia, and reports in the ADS 
typically contain a four letter word in the volume field such as ``conf",
``proc", ``book", ``coll", or ``rept".  When this is the case with a bibliographic 
code, the journal field typically consists of the first letter from 
important words in the title.  This can give the user the ability to 
identify a conference proceeding at a glance (e.g. ``ioda.book" for 
``Information and On-Line Data in Astronomy").  We will often leave the 
fifth place of the journal field as a dot for ``readability" 
(e.g. 1995ioda.book..175M).  For most
proceedings which are also published as part of a series (e.g. ASP Conference 
Series, IAU Colloquia, AIP Conference Series), we include in the system two
bibliographic codes, one as described above and one which contains the series 
name and the volume (see section \ref{masterlist}).  We do this so that users can see, for example, 
that a paper published in one of the ``Astronomical Data Analysis 
Software and Systems" series is clearly labelled as ``adass" whereas a typical
user might not remember which volume of ASPC contained those ADASS papers.
This increases the user's readability of bibliographic codes.

     With the STI data, the details were often unclear as to whether an
article was from a conference proceeding, a meeting, a colloquium, etc.
We assigned those codes as best we could, making no significant distinction
between them.  For conference abstracts submitted by the editors of 
a proceedings prior to publication, we often do not have page numbers.  
In this case, 
we use a counter in lieu of a page number and use an ``E" (for 
``Electronic") in the fourteenth column, the qualifier for publication.  
If these conference abstracts are then published, their bibliographic codes 
are replaced by a bibliographic code complete with page number.  
If the conference abstracts are published only on-line, they retain their 
electronic bibliographic code with its E and counter number.

     There are several other instances of datasets where the bibliographic
codes are non-standard.  PhD theses in the system use ``PhDT" as the
journal abbreviation, contain no volume number, and contain a counter in
lieu of a page number.  Since PhD theses, like all bibliographic codes,
are unique across all of the databases, the counter makes the bibliographic
code an identifier for only one thesis.  IAU Circulars also use a
counter instead of a page number.  Current Circulars are electronic in form,
and although not technically a new page, the second item of an IAU
Circular is the electronic equivalent of a second page.  Using the page 
number as a counter enables us to minimize use of the M identifier in
the fourteenth place of a bibliographic code for unduplicating.  This is desirable
since codes containing those identifiers are essentially impossible to 
create a priori, either by the journals or by users.

     The last set of data currently included in the ADS which contain non-standard
bibliographic codes is the ``QB"
book entries from the Library of Congress.  QB is the Library of Congress 
code for astronomy-related books and we have put approximately 17,000 of
these references in the system.  Because the QB numbers are identifiers by
themselves, we have made an exception to the bibliographic code format to use
the QB number (complete with any series or part numbers), prepended with
the publication year as the bibliographic code.  Such an entry is easily 
identifiable as a book, and these codes enable users to locate the books in 
most libraries.

     It is worth noting that while the bibliographic code makes identification
simple for the vast majority of references in the system, we are aware of two
instances where the bibliographic definition breaks down.  The use of the
fourteenth column for a qualifier such as ``L" for {\it ApJ Letters} makes
it impossible to use that column for unduplicating.  Therefore, if there
are two errata on the same page with the same author initial, there is
no way to create unique bibliographic codes for them.  We are aware of
only one such instance in the 33 years of publication of {\it ApJ Letters}.
Second, with the electronic publishing of an increasing number of journals, 
the requirement of page numbers to locate articles becomes unnecessary.  The
journal {\it Physical Review D} is currently using 6-digit article identifiers
as page numbers.  Since the bibliographic code allows for page 
numbers not longer than 5 digits, we are currently converting these
6-digit identifiers to their 5-digit hexagesimal equivalent.  Both of these
anomalies indicate that over the next few years we will likely need to
alter the current bibliographic definition in order to allow consistent
identification of articles for all journals.

\subsection{Data Fields}

     The databases are set up such that some data fields are searchable and 
others are not.  The searchable fields (title, author, and text) are the 
bulk of the important data, and these fields are indexed so that a query 
to the database returns the maximum number of meaningful results.
(see \cite{aa}, hereafter ARCHITECTURE).
The text field is the union of the abstract, title, keywords, and 
comments.  Thus, if a user requests a particular word in the text
field, all papers are returned which contain that word in the
abstract {\bf OR} in the title {\bf OR} in the keywords {\bf OR} in the comments.   Appendix A shows version 1.0 of the Extensible Markup Language (XML, see 
\ref{dataformats}) Document Type Definition (DTD) for text files in the
ADS Abstract Service.   The DTD lists fields 
currently used or expected to be used in text files in the ADS 
(see section \ref{textfiles} for details on the text files).  We intend to 
reprocess the current journal and affiliation fields in order to extract some 
of these fields.
 
     Since STI ceased abstracting the journal literature, we decided to
make the keywords themselves no longer a searchable entity for the time 
being -- they are 
searchable only through the abstract text field.  STI used a different 
standard set of keywords from the AAS journals, who use a different set of 
keywords from AIP journals (e.g. {\it AJ} prior to 1998).  In addition, keywords from
a single journal such as the {\it Astrophysical Journal (ApJ)} have evolved over
the years so that early {\it ApJ} volume keywords are not consistent with
later volumes.  In order to build one coherent 
set of keywords, an equivalence or synonym table for these different keyword 
sets must be created.  We are investigating different schemes for 
doing this, and currently plan to have a searchable keyword field again,
which encompasses all keywords in the system and equates those from different 
keyword systems which are similar (\cite{lee}).  

     The current non-searchable fields in the ADS databases include the 
journal field, author affiliation, category, abstract copyright, 
and abstract origin.  Although we may decide to create an index and
search interface for some of these entities (such as category), others 
will continue to remain unsearchable since searching them is not useful to the 
typical user.  In particular, author affiliations would be useful to search, 
however this information is inconsistently formatted so it is virtually 
impossible to collect all variations of a given institution for indexing 
coherently.  Furthermore, we have the author affiliations for only about 
half of the entries in the Astronomy database so we have decided to keep this 
field non-searchable.  For researchers wishing 
to analyze affiliations on a large scale, we can provide this information on 
a collaborative basis.

\subsection{Data Sources}

     The ADS currently receives abstracts or table of contents (ToC) references
from almost two hundred journal sources.  Tables~\ref{table2}, \ref{table3},
and \ref{table4} list these
journals, along with their bibliographic code abbreviation, source, frequency
with which we receive the data, what data are received, and any links we
can create to the data.  ToC references typically 
contain only author and title, although sometimes keywords are included as well.
The data are contributed via email, ftp, or retrieved from web sites around the 
world at a frequency ranging from once a week to approximately once a 
year.  The term ``often" used in the frequency column implies that we get
them more frequently than once a month, but not necessarily on a regular
basis.  The term ``occasionally" is used for those journals who submit data
to us infrequently.  

\begin{table*}
\caption[]{The ADS Astronomy Database }
\label{table2}
\begin{tabular*}{7.5in}{llp{3in}lll}

\hline
\noalign{\smallskip}

Journal & Source & Full Name & How Often & Kind of Data & Links$^{\mathrm{a}}$

\\
\noalign{\smallskip}
\hline
\noalign{\smallskip}
See accompanying text file ADS\_dataT2.txt for Table 2.
\\
\noalign{\smallskip}
\hline
\end{tabular*}
\begin{list}{}{}
\item[$^{\mathrm{a}}$]Letter codes describing what data are available
\item[$^{\mathrm{b}}$]Astronomische Gesellschaft
\item[$^{\mathrm{c}}$]University of Chicago Press
\item[$^{\mathrm{d}}$]American Institute of Physics
\item[$^{\mathrm{e}}$]Overseas Publishers Association
\item[$^{\mathrm{f}}$]American Geophysical Union
\item[$^{\mathrm{g}}$]Central Bureau for Astronomical Telegrams
\item[$^{\mathrm{h}}$]Academic Press
\item[$^{\mathrm{i}}$]Universitad Nacional Autonoma de Mexico
\item[$^{\mathrm{j}}$]Astronomisches Rechen-Institut
\end{list}{}{}
\end{table*}

\begin{table*}
\caption[]{The ADS Instrumentation Database }
\label{table3}
\begin{tabular*}{7.5in}{llp{3in}lll}

\hline
\noalign{\smallskip}
Journal & Source & Full Name & How Often & Kind of Data & Links$^{\mathrm{a}}$ \\
\noalign{\smallskip}
\hline
\noalign{\smallskip}
See accompanying text file ADS\_dataT3.txt for Table 3.
\\
\noalign{\smallskip}
\hline
\end{tabular*}
\begin{list}{}{}
\item[$^{\mathrm{a}}$]Letter codes describing what data are available
\item[$^{\mathrm{b}}$]Optical Society of America
\item[$^{\mathrm{c}}$]The International Society for Optical Engineering (SPIE)
\item[$^{\mathrm{d}}$]Institute of Physics
\end{list}{}{}
\end{table*}

\begin{table*}
\caption[]{The ADS Physics Database }
\label{table4}
\begin{supertabular*}{7.5in}{llp{3in}lll}

\hline
\noalign{\smallskip}
Journal & Source & Full Name & How Often & Kind of Data & Links$^{\mathrm{a}}$
\\
\noalign{\smallskip}
\hline
\noalign{\smallskip}
See accompanying text file ADS\_dataT4.txt for Table 4.
\\
\noalign{\smallskip}
\hline
\end{supertabular*}
\begin{list}{}{}
\item[$^{\mathrm{a}}$]Letter codes describing what data are available
\end{list}{}{}
\end{table*}

Updates to the Astronomy and Instrumentation databases occur approximately 
every two weeks, or more often if logistically possible, in order to keep 
the database current.  
Recent enhancements to the indexing software have enabled us to perform 
instantaneous updates, triggered by an email containing new data (see 
ARCHITECTURE).  Updates to the Physics database occurs approximately once 
every two months.  As stated earlier, the Preprint database is updated nightly.

\subsection{\label {dataformats} Data Formats}

     The ADS is able to benefit from certain standards which are adhered to in
the writing and submission practices of astronomical literature.  The journals
share common abbreviations and text formatting routines which are used by
the astronomers as well.  The use of TeX (\cite{knuth}) and LaTeX (\cite{lamport}), and their extension to 
BibTeX (\cite{lamport}) and AASTeX (\cite{aas}) results in common formats among 
some of our data sources.  This enables the reuse of parsing routines 
to convert these formats to our standard format.   Other variations of TeX 
used by journal publishers also allows us to use common parsing routines
which greatly facilitates data loading.

     TeX is a public domain typesetting program designed especially for math and
science.  It is a markup system, which means that formatting commands are 
interspersed with the text in the TeX input file. In addition to commands for 
formatting ordinary text, TeX includes many special symbols and commands with 
which you can format mathematical formulae with both ease and precision.
Because of its extraordinary capabilities, TeX has become the leading 
typesetting system for science, mathematics, and engineering. It was developed 
by Donald Knuth at Stanford University.

     LaTeX is a simplified document preparation system built on TeX.  Because 
LaTeX is available for just about any type of computer and because 
LaTeX files are ASCII, scientists are able to send their papers electronically 
to colleagues around the world in the form of LaTeX input.   This is also
true for other variants of TeX, although the astronomical publishing community
has largely centered their publishing standards on LaTeX or one of the software
packages based on LaTeX, such as BibTeX or AASTeX.  BibTeX is a 
program and file format 
designed by Oren Patashnik and Leslie Lamport in 1985 for the LaTeX document 
preparation system, and AASTeX is a LaTeX-based package that can be used to 
mark up manuscripts specifically for American Astronomical Society (AAS) journals.  

     Similar to the widespread acceptance of TeX and its variants, the extensive
use of SGML (Standard Generalized Markup Language, \cite{goldfarb}) by the members of the
publishing community has given us the ability to standardize many of our
parsing routines.  All data gleaned off the World Wide Web share features 
due to the use of HTML (HyperText Markup Language, \cite{powell}), an example of 
SGML.  Furthermore, the trend towards using XML (Extensible Markup Language, 
\cite{harold})
to describe text documents will enable us to share standard document
attributes with other members of the astronomical community.  
XML is a subset of SGML which is intended to enable generic SGML to be
served, received, and processed on the Web in the way that is now possible
with HTML.
The ADS parsing routines benefit from these standards in several ways:  
we can reuse routines designed around these systems; we are able to preserve 
original text representations of entities such as embedded accents so these 
entities are displayed correctly in the user's browser; and we are able to capture value-added 
features such as electronic URLs and email addresses for use elsewhere in 
our system.

     In order to facilitate data exchange between different parts of the ADS,
we make use of a tagged format similar to the ``Refer" format 
(\cite{jacobs}).
Refer is a preprocessor for the word processors nroff and troff which finds and
formats references.  While our tagged formats share some common fields 
(\%A, \%T, \%J, \%D), the Refer format is not specific enough to be used for
our purposes.  Items such as objects, URLs and copyright notices are beyond
the scope of the Refer syntax.  Details on our tagged format are provided in 
Table~\ref{table5}.  Reading and writing routines for this format are shared by 
loading and indexing routines, and a number of our data sources 
submit abstracts to us in this format.

\begin{table}
\caption[]{Tagged Format Definitions }
\label{table5}
\begin{tabular*}{3.4in}{lll}

\hline
\noalign{\smallskip}

Tag & Name & Comment

\\
\noalign{\smallskip}
\hline
\noalign{\smallskip}
\%R&Bibliographic Code         &required \\
\%T&Title                      &required \\
\%A&Author List                &required \\
\%D&Publication Date           &required \\
\%B&Abstract Text     & \\
\%C&Abstract Copyright   & \\
\%E&URL for Electronic Data Table     & \\
\%F&Author Affiliation     & \\
\%G&Origin & \\
\%H&Email & \\
\%J&Journal Name, Volume, and Page Range  & \\
\%K&Keywords     & \\
\%L&Last Page of Article     & \\
\%O&Object Name     & \\
\%Q&Category & \\
\%U&URL for Electronic Document    & \\ 
\%V&Language & \\
\%W&Database (AST, PHY, INST) & \\
\%X&Comment & \\
\%Y&Identifiers & \\
\%Z&References & \\

\\
\noalign{\smallskip}
\hline
\end{tabular*}
\end{table}

\section {\label {creating} Creating the Bibliographic Records}

One of the basic principles in the parsing and
formatting of the bibliographic data incorporated into the ADS database
over the years has been to preserve as much of the original information as 
possible and delay any syntactic or semantic interpretation of the
data until a later stage.  From the implementation point of view, this
means that bibliographic records provided to the ADS by publishers or
other data sources typically are saved as files which are tagged
with their origin, entry date, and any other ancillary information
relevant to their contents (e.g. if the fields in the record contain
data which was transliterated or converted to ASCII).

For instance, the records provided to the ADS by the University of
Chicago Press (the publisher of several major U.S. astronomical journals)
are SGML documents which contain a unique manuscript identifier
assigned to the paper during the electronic publishing process. 
This identifier is saved in the file created by the ADS system for 
this bibliographic entry.

Because data about a particular bibliographic
entry may be provided to the ADS by different sources and at different
times, we adopted a multi-step procedure in the creation and management
of bibliographic records:

  1) Tokenization: Parsing input data into a memory-resident data structure using procedures which are format- and source-specific. 

  2) Identification: Computing the unique bibliographic record identifier used by the ADS to refer to this record. 

  3) Instantiation: Creating a new record for each bibliography formatted according to the ADS ``standard" format. 

  4) Extraction: Selecting the best information from the different records available for the same bibliography and merging them into a single entry, avoiding duplication of redundant information. 

\subsection{Tokenization}

The activity of parsing a (possibly) loosely-structured bibliographic
record is typically more of an art than a science, given the wide
range of possible formats used by people for the representation and 
display of these records.
The ADS uses the PERL language (Practical Extraction and Report Language, \cite{wall91})
for implementing most of the routines associated with handling the data.  PERL is an
interpreted programming language optimized for scanning and 
processing textual data.  It was chosen over other programming
languages because of its speed and flexibility in handling text strings.  
Features such as pattern matching and regular expression substitution 
greatly facilitate manipulating the data fields.
To maximize flexibility in the parsing and formatting operations of
different fields, we have written a set of PERL library modules and
scripts capable of performing a few common tasks.
Some that we consider worth mentioning from the methodological point 
of view are listed below.  

\begin{itemize}
\item Character set conversion: electronic data are often delivered
to us in different character set encodings, requiring translation
of the data stream in one of the standard character sets expected by
our input scripts.  The default character set that has been
used by the ADS until recently is ``Latin-1'' encoding
(ISO-8859-1, \cite{iso}).  We are now in the process of converting to 
the use of Unicode characters (\cite{unicode}) encoded in UTF-8 
(UCS Transformation Format, 8--bit form).
The advantage of using Unicode is its universality (all character sets
can be mapped to Unicode without loss of information).  The advantage
of adopting UTF-8 over other encodings is mainly the software support
currently available (most of the modern software packages can already 
handle UTF-8 internally).  The adoption of Unicode and UTF-8 also
works well with our adoption of XML as the standard format for
bibliographic data.

\item Macro and entity expansion: Several of the highly structured document
formats in use today rely on the strengths of the formatting
language for the specification of some common formatting tasks or data
tokens.  Typically this means that LaTeX documents that are supplied
to us make use of one or more macro packages to perform some of the
formatting tasks.  Similarly, SGML documents will conform to some
Document Type Definition (DTD) provided to us by the publisher, and will make use of some
standard set of SGML entities to encode the document at the required
level of abstraction.  What this means for us is that even if most
of the input data comes to us in one of two basic formats
(TeX/LaTeX/BibTeX or SGML/HTML/XML), we must be able to parse a large 
number of document classes, each one defined by a different
and ever increasing set of specifications, be it a macro package or
a DTD. 
      
\item Author name formatting: Special care has been taken in parsing and
formatting author names from a variety of possible input formats
to the standard one used by the ADS.  The proper handling of author names is 
crucial to the integrity of the data in the ADS.  Without proper author 
handling, users would be unable to get complete listings on searches by 
author names which comprise approximately two-thirds of all searches
(see \cite{gei}, hereafter SEARCH).

Since the majority of our data sources do not provide author names
in our standard format (last name, first name or initial), our loading 
routines need to be able to invert
author names accurately, handling cases such as multiple word last names 
(Da Costa, van der Bout, Little Marenin) and suffixes (Jr., Sr., III).  
Any titles in an author's name (Dr., Rev.) were previously omitted, but are
now being retained in the new XML formatting of text files.

     The assessment of what constitutes a multiple word last name as 
opposed to a middle name is non-trivial since some names, such as Davis,
can be a first name (Davis Hartman), a middle name (A. G. Davis Philip),
a last name (Robert Davis), or some combination (Davis S. Davis).
Another example is how to determine when the name ``Van" is a first name
(Van Nguyen), a middle name (W. Van Dyke Dixon), or part of a last
name (J. van Allen).  Handling all of these cases correctly requires not
only familiarity with naming conventions worldwide, but an intimate 
familiarity with the names of astronomers who publish in the field.  We
are continually amassing the latter as we incorporate increasing amounts
of data into the system, and as we get feedback from our users.

\item Spell checking: Since many of the historical records entered in the
ADS have been generated by typesetting tables of contents, typographical
errors can often be flagged in an automated way using spell-checking
software.   We have developed a PERL software driver for the international 
ispell program, a UNIX utility, which can be used as a spell-checking 
filter on all input to be considered textual information.  A custom dictionary 
containing terms specific to astronomy and space sciences is used 
to increase the recognition capabilities of the software module.
Any corrections suggested by the spell-checker module are
reviewed by a human before the data are actually updated.

\item Language recognition: Extending the capability of the spell-checker,
we have implemented a software module which attempts to guess the language
of an input text buffer based on the percentage of words that it can
recognize in one of several languages: English, German, French, 
Spanish, or Italian.  This module is used to flag records to be
entered in our database in a language other than English.  Knowledge of the
language of an abstract allows us to create accurate synonyms for those 
words (see ARCHITECTURE).
\end{itemize}

\subsection{Identification}

We call identification the activity of mapping the tokens extracted from the 
parsing of a bibliographic record into a unique identifier.
The ADS adopted the use of bibliographic codes as the identifier for 
bibliographic entries shortly after its inception, in order to facilitate 
communication between the ADS and SIMBAD.  The advantage of using 
bibliographic codes as unique identifiers is
that they can most often be created in a straightforward way from the
information given in the list of references published in the
astronomical literature, namely the publication year, journal name,
volume, and page numbers, and first author's name (see section \ref{bibcodes} for details).

\subsection{Instantiation}

``Instantiation" of a bibliographic entry consists of the creation of a
record for it in the ADS database.  
The ADS must handle receipt of the same data from multiple sources.  We have
created a hierarchy of data sources so that we always know the preferred data
source.  A reference for which we have received records from STI, the journal publisher, SIMBAD, and NED, for 
example, must be in the system only once with the best information from each
source preserved.  When we load a reference into the system, we check whether 
a text file already exists for that reference.  If there is no text file, it
is a new reference and a text file is created.  If there already is a text file,
we append the new information to the current text file, creating a ``merged" 
text file.  This merged text file lists every instance of every field that we 
have received.  

\subsection{Extraction}

By ``extraction" of a bibliographic entry we mean the procedure
used to create a unique representation of the bibliography from the
available records.  This is essentially an activity of data fusion
and unification, which removes redundancies in the bibliographic
records obtained by the ADS and properly labels fields by their characteristics.
The extraction algorithm has been designed with our prior experience as 
to the quality of the data to select the best fields from each data 
source, to cross-correlate the fields as necessary, and to create a 
``canonical" text file which contains a unique instance of each field.
Since the latter is created through software, only one version of the text 
file must be maintained; when the merged text file is appended, the 
canonical text file is automatically recreated.

     The extraction routine selects the best pieces of information from each
source and combines them into one reference which is more complete than
the individual references.  For example, author lists received
from STI were often truncated after five or ten authors.  Whenever we
have a longer author list from another source, that author list is used
instead.  This not only recaptures missing authors, it also provides full
author names instead of author initials whenever possible.  In addition,
our journal sources sometimes omit the last page number of the
reference, but SIMBAD usually includes it, so we are able to preserve this
information in our canonical text file.

     Some fields need to be labelled by their characteristics so that they are
properly indexed and displayed.  The keywords, for example, need to be 
attributed to a specific keyword system.  The system designation allows for
multiple keyword sets to be displayed (e.g. NASA/STI Keywords and AAS Keywords)
and will be used in the keyword synonym table currently under development (\cite{lee}).

     We also attempt to cross-correlate authors with their affiliations
wherever possible.  This is necessary for records where the preferred author
field is from one source and the affiliations are from another source.
We attempt to assign the proper affiliation based on the last name and do not
assume that the author order is accurate since we are aware of ordering 
discrepancies in some of the STI records.

     Through these four steps in the procedure of creating and managing
bibliographic records, we are able to take advantage of receiving the 
same reference from multiple sources.  We standardize the various
records and present to the user a combination of the 
most reliable fields from each data source in one succinct text file.

\section {\label {updating} Updating the Database}

The software to update bibliographic records in the database consists 
of a series of PERL scripts, typically one 
per data source, which reads in the data, performs any 
special processing particular to that data source, and writes out the data
to text files.  The loading routines perform three fundamental tasks:  1) they 
add new
bibliographic codes to the current master list of bibliographic codes in 
the system; 2) they create and organize the text files containing the 
reference data; and 3) they maintain the lists of bibliographic codes 
used to indicate what items are available for a given reference.

\subsection{\label {masterlist} The Master List}

     The master list is a table containing bibliographic codes together with their
publication dates (YYYYMM) and entry dates into the system (YYYYMMDD).  There is 
one master list per database with one line per reference.  The most important
aspect of the master list is that it retains information about ``alternative"
bibliographic codes and matches them to their corresponding preferred
bibliographic code.  An alternative bibliographic code is usually a 
reference which we receive from another source (primarily SIMBAD or NED) which 
has been assigned a different bibliographic code from the one used by 
the ADS.  Sometimes this is due to the different rules used to build 
bibliographic codes for non-standard publications (see section \ref{bibcodes}), but often
it is just an incorrect year, volume, page, or author initial in one of the
databases (SIMBAD or NED or the ADS).  In either case, the ADS must keep 
the alternative bibliographic code in the system so that it can be found when 
referenced by the other source (e.g. when SIMBAD sends back a list of their 
codes related to an object).  The ADS matches the alternative bibliographic code 
to our corresponding one and replaces any instances of the alternative code 
when referenced by the other data source.  Alternative bibliographic codes in 
the master list are prepended with an identification letter (S for SIMBAD, 
N for NED, J for Journal) so that their origin is retained.
 
     While we make every effort to propagate corrections back to our data 
sources, sometimes there is simply a valid discrepancy.  For example, 
alternative bibliographic codes are often different from the ADS bibliographic code due to 
ambiguous differences such as which name is the surname of a Chinese author.  
Since Americans tend to invert Chinese names one way (Zheng, Wei) and Europeans 
another (Wei, Zheng), this results in two different, but equally valid codes.
Similarly, discrepancies in journal names such as BAAS (for the published
abstracts in the {\it Bulletin of the American Astronomical Society}) and AAS
(for the equivalent abstract with meeting and session number, but
no volume or page number) need different codes to refer to the same paper.
Russian and Chinese translation journals ({\it Astronomicheskii Zhurnal} vs.
{\it Soviet Astronomy} and {\it Acta Astronomica Sinica} vs. {\it Chinese Astronomy and
Astrophysics}) share the same problem.  These papers appear once in the
foreign journal and once in the translation journal (usually with different
page numbers), but are actually the same paper which should be in the system
only once.  The ADS must therefore maintain
multiple bibliographic codes for the same article since each journal 
has its own abbreviation, and queries for either one must be able to be
recognized.  The master list is the source of this correlation
and enables the indexing procedures and search engine to recognize 
alternative bibliographic codes.

\subsection{\label {textfiles} The Text Files}

     Text files in the ADS are stored in a directory tree by bibliographic code.
The top level of directories is divided into directories with four-digit names 
by publication year (characters 1 through 4 of the bibliographic code).  The
next level contains directories with five-character names according to 
journal (characters 5 through 9), and the text files are named by full 
bibliographic code under these journal directories.  Thus, a sample pathname is 
1998/MNRAS/1998MNRAS.295...75E.  Alternative bibliographic codes do not 
have a text file named by that code, since the translation to the equivalent 
preferred bibliographic code is done prior to accessing the text file.

A sample text file is given in the appendices.  Appendix B shows
the full bibliographic entry, including all records as received from 
STI, {\it MNRAS}, and SIMBAD.  It contains XML-tagged fields from each source, 
showing all instances of every field.  Appendix C shows the extracted
canonical version of the bibliographic entry which contains only selected 
information from the merged text file.  This latter version is displayed 
to the user through the user interface (see SEARCH).

\subsection{The Codes Files}

     The third basic function of the loading procedures is to modify and
maintain the listings for available items.  The ADS displays the availability
of resources or information related to bibliographic entries as letter codes in the
results list of queries and as more descriptive hyperlinks in the page
displaying the full information available for a bibliographic entry.
A full listing of the available item codes and their meaning is
given in SEARCH.

The loading routines maintain lists of bibliographic codes
for each letter code in the system which are converted to URLs 
by the indexing routines (see ARCHITECTURE).  Bibliographic codes are
appended to the lists either
during the loading process or as post-processing work depending on
the availability of the resource.   When electronic availability of data
coincides with our receipt of the data, the bibliographic codes can be
appended to the lists by the loading procedures.  When we receive the 
data prior to electronic availability, post-processing routines must be run 
to update the bibliographic code lists after we are notified that we 
may activate the links.

\section {\label {articles} The Articles}

     The ADS is able to scan and provide free access to past issues of the astronomical
journals because of the willing collaboration of the journal publishers.  The 
primary reason that the journal publishers have agreed to allow the scanning of 
their old volumes is that the loss of individual subscriptions does not pose
a threat to their livelihood.  Unlike many disciplines, most astronomy 
journals are able
to pay for their publications through the cost of page charges to astronomers
who write the articles and through library subscriptions which are unlikely
to be cancelled in spite of free access to older volumes through the ADS.  
The journal publishers continue to charge for access to the current volumes,
which is paid for by most institutional libraries.
This arrangement places astronomers in a fortunate position
for electronic accessibility of astronomy articles.

     The original electronic publishing plans for the astronomical community 
called for STELAR (STudy of Electronic Literature for Astronomical Research, 
\cite{vansteen92}, \cite{vansteen92al}, \cite{warnock92}, 
\cite{1993adass...2..137W})
to handle the scanning and dissemination of the full journal articles.
However, when the STELAR project was terminated in 1993, the ADS assumed 
responsibility for providing scanned full journal articles to the astronomical 
community.  The first test journal to be scanned was the {\it ApJ Letters} which was 
scanned in January, 1995 at 300 dots per inch (dpi).  It should be noted
that those scans were intended to be 600 dpi and we will soon rescan them
at the higher 600 dpi resolution.  Complications in the journal 
publishing format (plates at the end 
of some volumes and in the middle of others) were noted and detailed 
instructions provided to the scanning company so that the resulting scans 
would be named properly by page or plate number.

     All of the scans since the original test batch have been scanned at 600 dpi 
using a high speed scanner and generating a 1 bit/pixel monochrome image for 
each page.  The files created are then automatically processed in order to 
de-skew and center the text in each page, resize images to a standard U.S. 
Letter size (8.5 x 11 inches), and add a copyright notice at the bottom of 
each page.  For each original scanned page, two separate image files of 
different resolutions are generated and stored on disk.  The availability of 
different resolutions allows users the flexibility of downloading either high 
or medium quality documents, depending on the speed of their internet 
connection.  The image formats and compression used were 
chosen based on the available compression algorithms and browser capabilities.
The high resolution files currently used are 600 dpi, 1 bit/pixel TIFF 
(Tagged Image File Format) files,
compressed using the CCITT Group 4 facsimile 
encoding algorithm.  The medium resolution files are 200 dpi, 1 bit/pixel 
TIFF files, also with CCITT Group 4 
facsimile compression.  

     Conversion to printing formats (PDF, PCL, and Postscript) is done on demand,
as requested by the user.  Similarly, conversion from the TIFF files to 
a low resolution GIF (Graphic Interchange Format) file 
(75, 100, or 150 dpi, depending on user preferences)
for viewing on the computer screen is done on demand, then cached so that 
the most frequently accessed pages do not need to be created every time.  
A procedure run nightly deletes the GIF files with the oldest access time stamp 
so that the total size of the disk cache is kept under a pre-defined limit.
The current 10 GBytes of cache size in use at the SAO Article Server causes
only files which have not been accessed for about a month to be deleted.
Like the full-screen GIF images, the ADS also caches thumbnail images of
the article pages which provide users with the capability of viewing 
the entire article at a glance.

     The ADS uses Optical Character Recognition (OCR) software to gain
additional data from TIFF files of article scans.
The OCR software is not yet adequate
for accurate reproduction of the scanned pages.  Greek symbols, equations, 
charts, and tables do not translate accurately enough to remain true to the 
original printed page.   For this reason, we have chosen not to
display to the user anything rendered by the OCR software in an unsupervised
fashion.  However, we are 
still able to take advantage of the OCR software for several purposes.

     First, we are able to identify and extract the abstract paragraph(s) for
use when we do not have the abstract from another source.   In these cases, 
the OCR'd text is indexed so that it is searchable and the extracted image of the 
abstract paragraph is displayed in lieu of an ASCII version of the abstract.  
Extracting the abstract from the scanned pages is somewhat tedious, as it requires establishing different sets 
of parameters for each journal, as well as for different fonts used over the 
years by the same journal.  The OCR software can be taught how to determine 
where the abstract ends, but it does not work for every article due 
to oddities such as author lists which extend beyond the first page of an 
article, and articles which are in a different format from others in the 
same volume (e.g. no keywords or multiple columns).  The ADS currently
contains approximately 25,000 of these abstract images and more will be added 
as we continue to scan the historical literature.

     We are also currently using the OCR software to render electronic 
versions of the entire scanned articles for indexing purposes.  We 
will not use this for display to the users, but hope to be able to index
it to provide the possibility of full text searching at some future date.
We estimate that the indexing of our almost one million 
scanned pages with our current hardware and software will take 
approximately two years of dedicated CPU time.

     The last benefit that we gain from the OCR software is the
conversion of the reference list at the end of articles.  
We use parsed reference lists from the scanned articles 
to build citation and reference lists for display through the C and 
R links of the available items.  Since reference lists are typically in one of
several standard formats, we parse each reference for author, journal, 
volume and page number for most journal articles, and conference name, author, and
page number for many
conference proceedings.  This enables us to build
bibliographic code lists for references contained in that article (R links)
and invert these lists to build bibliographic code lists of articles 
which cite this paper (C links).  We are able to use this process to identify
and therefore add commonly-cited articles which are currently missing from
the ADS.  This is usually data prior to 1975 or astronomy-related articles
published in non-astronomy journals.

The Article Service currently contains 250 GBytes of scans, which
consists of 1,128,955 article pages comprising 138,789 articles.  These
numbers increase on a regular basis, both as we add more articles from the
older literature and as we scan new journals.

\section {\label {journals} ADS/Journal Interaction}

A description of the data in the ADS would be incomplete without a 
discussion of the interaction between the ADS and the electronic 
journals.  The data available on-line from the journal publishers is an extension
of the data in the ADS and vice versa.  This interaction is greatly
facilitated by the acceptance of the bibliographic code by many
journal publishers as a means for accessing their on-line articles.

Access to articles currently on-line at the journal sites through the 
ADS comprises a significant percent of the on-line journal access (see
OVERVIEW).  The best model for interaction between the ADS and a journal
publisher is the University of Chicago Press (hereafter UCP), publisher of
{\it ApJ, ApJL, ApJS, AJ,} and {\it PASP}.  When a new volume appears on-line at
UCP, the ADS is notified by email and an SGML header file for each of those
articles is simultaneously transferred to our site.  The data are parsed
and loaded into the system and appropriate links are created.  However,
prior to this, the UCP has made use of the ADS to build their electronic
version through the use of our bibliographic code reference resolver.

     Our bibliographic code reference resolver (\cite{adass8}) was developed to provide
the capability to automatically parse, identify, and verify citations
appearing in astronomical literature.  By verifying the existence of a
reference through the ADS, journals and conference proceedings editors are
able to publish documents containing hyperlinks pointing to stable, unique
URLs.  Increasingly more journals are linking to the ADS in their
reference sections, providing users with the ability to read referenced
articles with the click of a mouse button.

During the copy editing phase, UCP editors query the ADS reference
resolver and determine if each reference exactly matches a bibliographic
code in the ADS.  If there is a match, a link to the ADS is established for
this entry in their reference section.  If there is not a match, one of
several scenarios takes place.  First, if it is a valid reference not yet
included in the ADS (most often the case for ``fringe" articles, those 
peripherally associated with astronomy), our reference resolver captures
the information necessary to add it to our database during the next update.
Second, if it is a valid reference unable to be parsed by the resolver
(sometimes the case for conference proceedings or PhD theses), no 
action is taken and no link is listed in the reference section.  Third,
if there is an error in the reference as determined by the reference resolver,
the UCP editors may ask for a correction or clarification from the authors.

The last option demonstrates the power of the reference resolver, which has
been taught on a journal-by-journal basis how complete the coverage of
that journal is in the ADS.  Before the implementation of the reference resolver,
UCP was able to match 72\% of references in {\it ApJ} articles (E. Owens,
private communication).  Early results from the use of the reference 
resolver show that we are now able to match conference proceedings, so this
number should become somewhat larger.  It is unlikely that we will ever match
more than 90\% of references in an article
due to references such as ``private communication", ``in press", and preprints,
as well as author errors (see section \ref {summary}).  Our own reference 
resolving of OCR'd reference lists shows that we can match approximately
86

     The ADS provides multiple ways for authors and journal publishers to link 
to the ADS (see SEARCH).  We make every effort to 
facilitate individuals and organizations linking to us.  This is easily 
done for simple searches such as the verification of a bibliographic code or 
an author search for a single spelling.  However, given the complexity of 
the system, these automated searches can quickly become complicated.  
Details for conference proceedings editors or journal publishers who are
interested in establishing or improving links to the ADS are available upon 
request.  In particular, those who have individual TeX macros incorporated 
in their references can use our bibliographic code resolver to 
facilitate linking to the ADS.

\section {\label {summary} Discussion and Summary}

\begin{figure}
\resizebox{\hsize}{!}{\includegraphics{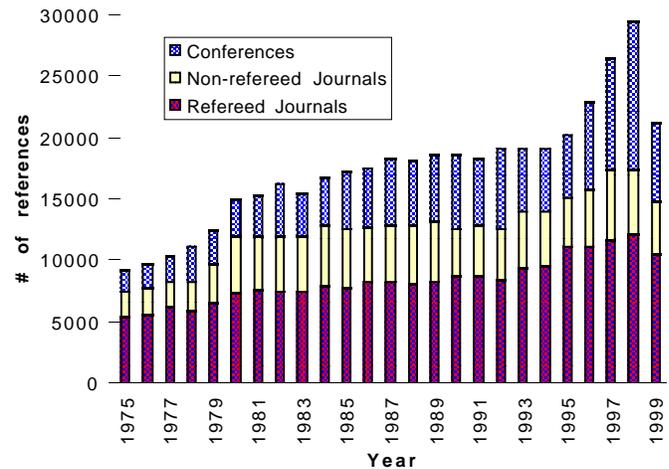}}
\caption[]{Histogram showing the number of refereed journal  papers, non-refereed journal papers, and conferences (including reports  and theses) for each year in the Astronomy database since 1975. }
\label{histogram}
\end{figure}

     As of this writing (12/1999), there are 524,304 references in the 
Astronomy database, 523,498 references in the Instrumentation database, 443,858
references in the Physics database, and 3467 references in the Preprint
database, for a total of almost 1.5 million references in the system.  
Astronomers currently write approximately 18,000 journal articles annually,
and possibly that many additional conference proceedings papers per year.  
More than half of the journal papers appear in peer-reviewed journals.  These 
numbers are more than double what they were in 1975, in spite of an increase
in the number of words per page in most of the major journals 
(\cite{1995ApJ...455..407A}),
and an increase in number of pages per article (\cite{1997PASP..109.1278S}).
At the current rate of publication, astronomers could be writing
25,000 journal papers per year by 2001 and an additional 20,000 conference proceedings papers.
Figure \ref{histogram} shows the total number of papers for each year in the
Astronomy database since 1975, divided into refereed journal papers,
non-refereed journal papers, and conferences (including reports and theses).
There are
three features worth noting.  First, the increase in total references in 1980 
is due to the inclusion
of Helen Knudsen's Monthly Astronomy and Astrophysics Index, a rich source
of data for both journals and conference proceedings which began coverage
in late 1979 and continued until 1995.  Second, the recent
increase in conferences included in the Astronomy database (starting
around 1996) is due to the inclusion of conference proceedings table
of contents provided by collaborating librarians and typed in by our 
contractors.  Last, the decrease in numbers for 1999 is due to coverage
for that year not yet being complete in the ADS.  

The growth rate of the Instrumentation and Physics databases is difficult 
to estimate, primarily because we do not have datasets which are as complete
as astronomy.  In any case, the need for the organization and maintenance
of this large volume of data is clearly important to every research astronomer.
Fortunately, the ADS was designed to be able to handle this large quantity of
data and to be able to grow with new kinds of data. New available item links 
have been added for new types of data as they became available (e.g. the links
to complete book entries at the Library of Congress)
and future datasets (e.g. from future space missions) should be able to be 
added in the same fashion.  

     As with any dataset of this magnitude, there is some fraction of 
references in the system which are incorrect.  This is unavoidable given
the large number of data sources, errors in indices and tables of
contents as originally published, and human error.  In addition, many 
authors do not give full attention to verifying all references in a paper,
resulting in the introduction of errors in many places.  In a 
systematic study of more than 1000 references contained in a single issue of 
the {\it Astrophysical Journal},
Abt (1992) found that more than 12\% of those contained errors.  This number should be significantly reduced with the integration of the ADS reference
resolver in the electronic publishing process.
However, any mistakes in the ADS can and will get propagated, so steps are
being taken by us to maximize accuracy of our entries.

     Locating and identifying correlations between multiple bibliographic 
codes which describe the same article is a time-consuming and
sometimes subjective task as many pairs of bibliographic codes need to
be verified by manually looking up papers in the library.  We use the
Abstract Service itself for gross matching of bibliographic codes, submitting
a search with author and title, and considering any resulting matches with 
a score of 1.0 as a potential match.  These matches are only potential
matches which require verification since authors can submit the same paper
to more than one publication source (e.g. BAAS and a refereed journal), 
and since errata published with the same title and author list will perfectly
match the original paper.

When a volume or year is mismatched, it is usually obvious which of a pair
of matched bibliographic codes is correct, but if a page number is off,
the decision as to which code is correct cannot always be automated. We also
need to consider matches with very high scores less than 1.0 since these are the
matches where an author name may be incorrect.  The correction of errors
of this sort is ongoing work which is carried out as often as time and 
resources permit.

     The evolution of the Internet and the World Wide Web, along with the 
explosion of astronomical services on the Web has enabled the ADS to provide
access to our databases in an open and uniform environment.  We have been able
to hyperlink both to our own resources and to other on-line resources such
as the journal bibliographies (\cite{1996adass...5..547B}).   As part of the 
international collaboration Urania (Universal Research Archive of Networked 
Information in Astronomy, \cite{1998lisa.conf..107B}), the ADS enables
a fully functioning distributed digital library of astronomical 
information which provides power and utility previously unavailable to 
the researcher.  

     Perhaps the largest factor which has contributed to the success of the
ADS is the willing cooperation of the AAS, CDS, and all the journal publishers.
The ADS has largely become the means for linking together smaller pieces of a 
bigger picture, making an elaborate digital library for astronomers a reality.
We currently collaborate with over fifty groups in creating and maintaining
cross-links among data centers.  
These additional collaborations with individuals and institutions
worldwide allow us to provide many value-added features to the system such 
as object information, author email addresses, mail order forms for articles,
citations, article scans, and more.  A listing of these collaborations is 
provided in Table~\ref{table6}.  Any omissions from this table are purely
unintentional, as the ADS values all of our colleagues and the users benefit
not only from the major collaborators but the minor ones as well, as these are 
often more difficult for users to learn about independently.  Most of the
abbreviations are listed in Tables 2, 3, and 4.

\begin{table*}
\caption[]{Collaborators }
\label{table6}
\begin{tabular*}{7.0in}{lp{3.5in}}

\hline
\noalign{\smallskip}

Additional Collaborations & Nature of the Collaboration

\\
\noalign{\smallskip}
\hline
\noalign{\smallskip}
A.G. Davis Philip                            & Scanning of Conference Proceedings \\
Academic Press (AP)                          & Scanning of Icarus \\
American Astronomical Society (AAS)          & Citations, Scanning of AJ, ApJ, ApJL, ApJS, AASPB$^{\mathrm{a}}$, BAAS \\
American Institute of Physics                & Scanning of SvAL  \\
Andre Heck                                   & Star Heads (Author Home Pages) \\
Annual Reviews, Inc.                         & Scanning of ARA\&A \\
Astronomical Data Center (ADC)               & D links to data \\
Astronomical Institute of Czechoslovakia & Scanning of BAICz \\
Astronomical Institute of the Slovak Academy of Sciences & Scanning of CoSka \\
Astronomical Society of Australia  & Scanning of PASA \\
Astronomical Society of India                & Scanning of BASI \\
Astronomical Society of Japan   & Scanning of PASJ \\
Astronomical Society of the Pacific (ASP)    & Scanning of PASP and Conference Proceedings \\
Astronomische Gesellschaft                    & Scanning of RvMA \\
Astronomische Nachrichten                    & Scanning of AN \\
Baltic Astronomy                             & Scanning of BaltA \\
British Astronomical Association             & Scanning of JBAA \\
Cambridge University Press                   & M links to order forms, Scanning \\
Central Bureau for Astronomical Telegrams (CBAT)   & Object searches \\
Chris Benn                                   & Astropersons.lis (Author Email) \\
EDP Sciences                                 & Scanning of A\&AS \\
Elsevier Publishers                          & E links to articles \\
General Catalogue of Photometric Data (GCPD) & D links to data \\
Institute for Scientific Information (ISI)   & Citations \\
International Society for Optical Engineering (SPIE)& M links to order forms \\
Korean Astronomical Society                  & Scanning of JKAS \\
Kluwer Publishers                            & M links to order forms, Scanning of SoPh  \\
Library of Congress (LOC)                    & Z39.50 interface, L links to data \\
Los Alamos National Laboratory (LANL)        & Preprint Archive \\
Lunar and Planetary Science Institute (LPI)  & Scanning, Object searches \\
Meteoritical Society                         &  Scanning of M\&PS \\
NED                                          & N links to objects, Object searches \\
The Observatory                              & Scanning \\
Royal Astronomical Society                   & Scanning of MNRAS \\
SIMBAD                                       & S links to objects, D links to data, Object searches \\
Springer Verlag                          & Scanning of A\&A, ZA$^{\mathrm{b}}$ \\
Universitad Nacional Autonoma de Mexico (UNAM) & Scanning of RMxAA, RMxAC \\
University of Chicago Press (UCP)            & Reference Resolving \\

\\
\noalign{\smallskip}
\hline
\end{tabular*}
\begin{list}{}{}
\item[$^{\mathrm{a}}$]American Astronomical Society Photo Bulletin
\item[$^{\mathrm{b}}$]Zeitschrift f\"ur Astrophysik
\end{list}{}{}
\end{table*}

The successful coordination of data exchanges with each of our collaborators
and the efforts which went into establishing them in the first place have been 
key to the success of the ADS.  Establishing links to and from the journal 
publishers, changing these links due to revisions at publisher websites, and 
tracking and fixing broken links is all considered routine data maintenance
for the system.  Since it is necessary for us to maintain connectivity 
to external sites, routine checks of sample links are performed on a regular
basis to verify that the links are still active.

     Usage statistics for the Abstract Service (see OVERVIEW)
indicate that astronomers and librarians at scientific institutions
are eager to take advantage of the information that the ADS provides.  The 
widespread acceptance of the ADS by the astronomical community is changing how
astronomers do research, placing extensive bibliographic information at 
their fingertips.  This enables researchers to increase their productivity and
to improve the quality of their work.

     A number of improvements to the data in the ADS are planned for the near
future.  As always, we will continue our efforts to increase the completeness
of coverage, particularly for the data prior to 1975.  We have collected
most of the major journals back to the first issue for scanning and adding
to the Astronomy database.   In addition, we are scanning and OCR'ing table 
of contents for conference proceedings to improve our coverage in that
area.   We are currently OCR'ing full journal articles 
to provide full text searching and to improve the completeness of our 
reference and citation coverage.  Finally, as the ADS becomes commonplace for all 
astronomers, valuable feedback from our users to inform us about missing 
papers, errors in the database, and suggested improvements to the system 
serve to guide the future of the ADS and to ensure that the ADS continues to 
evolve into a more valuable research tool for the scientific community.

\section{\label{acknowldgements} Acknowledgments}

The other ADS Team members: Markus Demleitner, Elizabeth Bohlen, and Donna
Thompson contribute much on a daily basis.  Funding for this project has been
provided by NASA under NASA Grant NCC5-189.

\newpage

\appendix
\section {\label {appendixa}}

Version 1.0 of the XML DTD describing text files in the ADS Abstract
Service.

\begin{verbatim}
Document Type Definition for the ADS 
bibliographic records

Syntax policy
=============
 - The element names are in uppercase in order
   to help the reading.
 - The attribute names are preferably in 
   lowercase 
 - The attribute values are allowed to be of 
   type CDATA to allow more flexibility for 
   additional values; however, attributes
   typically may only assume one of a well-
   defined set of values
 - Cross-referencing among elements such as 
   AU, AF, and EM is accomplished through the 
   use of attributes of type IDREFS (for AU) 
   and ID (for AF and EM)

<!-- BIBRECORD is the root element of the XML 
     document.  Attributes are:

   origin  mnemonic indicating individual(s)
           or institution(s) who submitted 
           the record to ADS
   lang    language in which the contents of 
           this record are expressed the 
           possible values are language tags 
           as defined in RFC 1766.  
           Examples: lang="fr",  lang="en"
-->

<!ELEMENT BIBRECORD ( METADATA?, 
                      TITLE?, 
                      AUTHORS?, 
                      AFFILIATIONS?, 
                      EMAILS?, 
                      FOOTNOTES?, 
                      BIBCODE, 
                      MSTRING, 
                      MONOGRAPH?, 
                      SERIES?, 
                      PAGE?, 
                      LPAGE?, 
                      COPYRIGHT?, 
                      PUBDATE, 
                      CATEGORIES*, 
                      COMMENTS*, 
                      ANOTE?, 
                      BIBTYPE?, 
                      IDENTIFIERS?, 
                      ORIGINS, 
                      OBJECTS*, 
                      KEYWORDS*, 
                      ABSTRACT* ) >

<!ATTLIST BIBRECORD  origin CDATA   #REQUIRED
                     lang   CDATA   #IMPLIED  >

<!-- Generic metadata about the ADS record 
     (rather than the publication) -->
<!ELEMENT METADATA ( VERSION, 
                     CREATOR, 
                     CDATE, 
                     EDATE ) >

<!-- Versioning is introduced to allow parsers 
     to detect and reject any documents not 
     complying with the supported DTD  -->
<!ELEMENT VERSION ( #PCDATA ) >
<!-- CREATOR is purely informative -->
<!ELEMENT CREATOR ( #PCDATA ) >
<!-- Creation date for the record -->
<!ELEMENT CDATE ( YYYY-MM-DD ) >
<!-- Last modified date -->
<!ELEMENT EDATE ( YYYY-MM-DD ) >

<!-- Title of the publication -->
<!ELEMENT TITLE ( #PCDATA ) >
<!ATTLIST TITLE lang CDATA #IMPLIED >

<!-- AUTHORS contains only AU subelements, each 
     one of them corresponding to a single author 
     name -->
<!ELEMENT AUTHORS ( AU+ ) >

<!-- AU contains at least the person's last name 
     (LNAME), and possibly the first and middle name(s) 
     (or just the initials) which would be stored in 
     element FNAME.  PREF and SUFF represent the 
     salutation and suffix for the name.  SUFF 
     typically is one of: Jr., Sr., II, III, IV.  
     PREF is rarely used but is here for completeness.
     Typically we would store salutations such as 
     "Rev." (for "Reverend"), or "Prof." (for 
     "Professor") in this element. 
-->
<!ELEMENT AU ( PREF?, 
               FNAME?, 
               LNAME, 
               SUFF? ) >
<!-- The attributes AF and EM are used to cross-
		 reference author affiliations and email 
     addresses with the individual author records.  
     This is the only exception of attributes in 
     upper case.  The typical use of this is:
     <AU AF="AF_1 AF_2" EM="EM_3">...</AU>
-->
<!ATTLIST AU     AF     IDREFS  #IMPLIED
                 EM     IDREFS  #IMPLIED
                 FN     IDREFS  #IMPLIED >
<!-- AU subelements -->
<!ELEMENT PREF  ( #PCDATA ) >
<!ELEMENT FNAME ( #PCDATA ) >
<!ELEMENT LNAME ( #PCDATA ) >
<!ELEMENT SUFF  ( #PCDATA ) >

<!-- AFFILIATIONS is the wrapper element for 
     the individual affiliation records, each 
     represented as an AF element -->
<!ELEMENT AFFILIATIONS ( AF+ ) >
<!ELEMENT AF ( #PCDATA ) >
<!-- the value of the ident attribute should 
     match one of the values assumed by the AF 
     attribute in an AU element -->
<!ATTLIST AF         ident  ID      #REQUIRED >

<!ELEMENT EMAILS ( EM+ ) >
<!ELEMENT EM ( #PCDATA ) >
<!-- the value of the ident attribute should 
     match one of the values assumed by the EM 
     attribute in an AU element -->
<!ATTLIST EM         ident  ID      #REQUIRED >

<!-- FOOTNOTES and FN subelements are here for 
     future use -->
<!ELEMENT FOOTNOTES ( FN+ ) >
<!ELEMENT FN ( #PCDATA ) >
<!ATTLIST FN         ident  ID      #REQUIRED >

<!-- BIBCODE; for a definition, see:
http://adsdoc.harvard.edu/abs_doc/bib_help.html
http://adsabs.harvard.edu/cgi-bin/
       nph-bib_query?1995ioda.book..259S
http://adsabs.harvard.edu/cgi-bin/
       nph-bib_query?1995VA.....39R.272S
     This identifier logically belongs to the 
     IDENTS element, but since it is the 
     identifier used internally in the system, 
     it is important to have it in a prominent 
     and easy to reach place.
-->
<!ELEMENT BIBCODE ( #PCDATA ) >

<!-- MSTRING is the unformatted string for the 
    monograph (article, book, whatever).  Example:
    <MSTRING>The Astrophysical Journal, Vol. 526, 
     n. 2, pp. L89-L92</MSTRING>
-->
<!ELEMENT MSTRING ( #PCDATA ) >
<!-- MONOGRAPH is a structured record containing 
     the fielded information about the monograph 
     where the bibliographic entry appeared.  
     Typically this is created by parsing the 
     text in the MSTRING element.  Example:
      <MTITLE>The Astrophysical Journal</MTITLE>
      <VOLUME>526</VOLUME>
      <ISSUE>2</ISSUE>
      <PUBLISHER>University of Chicago Press
         </PUBLISHER>
-->
<!ELEMENT MONOGRAPH ( MTITLE, 
                      VOLUME?, 
                      ISSUE?, 
                      MNOTE?, 
                      EDITORS?, 
                      EDITION?, 
                      PUBLISHER?, 
                      LOCATION?, 
                      MID* ) >

<!-- Monograph title (e.g. "Astrophysical Journal") -->
<!ELEMENT MTITLE ( #PCDATA ) >
<!ELEMENT VOLUME ( #PCDATA ) >
<!ATTLIST VOLUME     type   NMTOKEN #IMPLIED  >
<!ELEMENT ISSUE ( #PCDATA ) >
<!-- A note about the monograph as supplied by the 
     publisher or editor -->
<!ELEMENT MNOTE ( #PCDATA ) >
<!-- List of editor names as extracted from MSTRING.
     Formatting is as for AUTHORS and AU elements -->
<!ELEMENT EDITORS ( ED+ ) >
<!ELEMENT ED ( PREF?, 
               FNAME?,
               LNAME,
               SUFF? ) >
<!-- Edition of publication -->
<!ELEMENT EDITION ( #PCDATA ) >
<!-- Name of publisher -->
<!ELEMENT PUBLISHER ( #PCDATA ) >
<!-- Place of publication -->
<!ELEMENT LOCATION ( #PCDATA ) >
<!-- MID represents the monograph identification as 
     supplied by the publisher.  This may be useful in 
     correlating our record with the publisher's online 
     offerings.  The "system" attribute characterizes 
     the system used to express the identifier -->
<!ELEMENT MID ( #PCDATA ) >
<!ATTLIST MID        type   NMTOKEN #IMPLIED  >

<!-- If the bibliographic entry appeared in a series, 
     then the element SERIES contains information 
     about the series itself.  Typically this consists 
     of data about a conference series (e.g. ASP 
     Conference Series).  Note that there may be 
     several SERIES elements, since some 
     publications belong to "subseries" within 
     a series.
-->
<!ELEMENT SERIES ( SERTITLE,
                   SERVOL?,
                   SEREDITORS?,
                   SERBIBCODE? ) >
<!-- Title, volume, and editors of conference 
     series -->
<!ELEMENT SERTITLE ( #PCDATA ) >
<!ELEMENT SERVOL ( #PCDATA ) >
<!ELEMENT SEREDITORS ( ED+ ) >
<!-- Serial bibcode for publication (may coincide 
     with main bibcode) -->
<!ELEMENT SERBIBCODE ( #PCDATA ) >

<!-- PAGE may have the attribute type set to 
     "s" for (sequential) the value associated 
     to it does not represent a printed volume 
     number -->
<!ELEMENT PAGE ( #PCDATA ) >
<!ATTLIST PAGE       type   NMTOKEN #IMPLIED  >

<!-- LPAGE gives the last page number (if known).
     Does not make sense if PAGE is type="s" -->
<!ELEMENT LPAGE ( #PCDATA ) >

<!-- COPYRIGHT is just an unformatted string 
     containing copyright information from 
     publisher -->
<!ELEMENT COPYRIGHT ( #PCDATA ) >

<!ELEMENT PUBDATE ( YEAR, MONTH? ) >
<!ELEMENT MONTH ( #PCDATA ) >
<!ELEMENT YEAR ( #PCDATA ) >

<!-- CATEGORIES contain subelements indicating in 
     which subject categories the publication was 
     assigned.  STI/RECON has always assigned a 
     category for each entry in their system, but 
     otherwise there is little else in our 
     database.  The attributes origin and system 
     are used to keep track of the different 
     classifications used.
-->
<!ELEMENT CATEGORIES ( CA+ ) >
<!ATTLIST CATEGORIES origin NMTOKEN #IMPLIED
                     system NMTOKEN #IMPLIED  >
<!ELEMENT CA ( #PCDATA ) >

<!-- Typically private fields supplied by the 
     data source.  For instance, SIMBAD and LOC 
     provide comments about a bibliographic 
     entries -->
<!ELEMENT COMMENTS ( CO+ ) >
<!ATTLIST COMMENTS lang   CDATA   #IMPLIED
                   origin NMTOKEN #IMPLIED >
<!ELEMENT CO ( #PCDATA ) >

<!-- Author note -->
<!ELEMENT ANOTE ( #PCDATA ) >

<!-- BIBTYPE describes what type of publication 
     this entry corresponds to.  This is 
     currently limited to the following tokens 
     (taken straight from the BibTeX 
     classification):
          article
          book
          booklet
          inbook
          incollection
          inproceedings
          manual
          masterthesis
          misc
          phdthesis
          proceedings
          techreport
          unpublished
-->
<!ELEMENT BIBTYPE ( #PCDATA ) >

<!-- List of all known identifiers for this 
     publication -->
<!ELEMENT IDENTIFIERS ( ID+ ) >
<!-- Contents of an ID element is the identifier 
     used by a particular publisher or institution.
     Examples:
       <ID origin="UCP" system="PUBID">38426</ID>
       <ID origin="STI" system="ACCNO">A90-12345</ID>
-->
<!ELEMENT ID ( #PCDATA ) >
<!ATTLIST ID         origin NMTOKEN #IMPLIED
                     type   NMTOKEN #REQUIRED >

<!-- the collective list of institutions that have given 
     us a record about this entry.  -->
<!ELEMENT ORIGINS ( OR+ ) >
<!ELEMENT OR ( #PCDATA ) >

<!-- The list of objects associated with the 
     publication -->
<!ELEMENT OBJECTS ( OB+ ) >
<!ELEMENT OB ( #PCDATA ) >

<!-- Keywords assigned to the publication -->
<!ELEMENT KEYWORDS ( KW+ ) >
<!ATTLIST KEYWORDS   Lang   CDATA   #IMPLIED
                     origin NMTOKEN #IMPLIED
                     system NMTOKEN #REQUIRED >
<!ELEMENT KW ( #PCDATA ) >

<!-- An abstract of the publication.  This is 
     typically provided to us by the publisher, 
     but may in some cases come from other 
     sources (E.g. STI, which keyed abstracts 
     in most cases).  Therefore we allow several 
     ABSTRACT elements within each record, each 
     with a separate origin or language.  
     The attribute type is used to keep track 
     of how the abstract data was generated.  
     For instance, abstract text generated by 
     our OCR software will have: 
         origin="ADS" type="OCR" lang="en"
-->
<!ELEMENT ABSTRACT ( P+ ) >
<!ATTLIST ABSTRACT   origin NMTOKEN #IMPLIED  >
                     type   NMTOKEN #IMPLIED  >
                     lang   CDATA   #IMPLIED  >

<!-- Abstracts are composed of separate 
     paragraphs which have mixed contents as 
     listed below.  All the subelements listed 
     below have the familiar HTML meaning and 
     are used to render the abstract text in a 
     decent way -->
<!ELEMENT P (#PCDATA |A| BR | PRE | SUP | SUB)* >
<!-- Line breaks (BR) and preformatted text (PRE) 
     make it possible to display tables and other 
     preformatted text. -->
<!ELEMENT BR EMPTY >
<!ELEMENT PRE  (#PCDATA | A | BR | SUP | SUB )* >
<!-- A is the familiar anchor element.  -->
<!ELEMENT A ( #PCDATA | BR | SUP | SUB )* >
<!ATTLIST A          HREF   CDATA   #REQUIRED >
<!-- SUP and SUB are superscripts and subscripts.  
     In our content model, they are allowed to 
     contain additional SUP and SUB elements, 
     although we may decide to restrict them to
     PCDATA at some point -->
<!ELEMENT SUP ( #PCDATA | A | BR | SUP | SUB )* >
<!ELEMENT SUB ( #PCDATA | A | BR | SUP | SUB )* >
\end{verbatim}

\section {\label {appendixb}}

A sample text file from the ADS Abstract Service showing 
XML markup for the full bibliographic entry, including records from
STI, {\it MNRAS}, and SIMBAD.  Items in bold are those selected to
create the canonical text file shown in Appendix C.

\noindent
$<$?xml version=``1.0"?$>$ \\
$<$!DOCTYPE ADS\_BIBALL SYSTEM ``ads.dtd"$>$ \\
$<$ADS\_BIBALL$>$ \\

\noindent
$<$BIBRECORD origin=``STI"$>$ \\
{\bf $<$TITLE$>$Spectroscopic confirmation of redshifts 
predicted by gravitational lensing$<$/TITLE$>$} \\
$<$AUTHORS$>$ \\
  $<$AU AF=``1"$>$ \\
    $<$FNAME$>$Tim$<$/FNAME$>$ \\
    $<$LNAME$>$Ebbels$<$/LNAME$>$ \\
  $<$/AU$>$ \\
  $<$AU AF=``1"$>$ \\
    $<$FNAME$>$Richard$<$/FNAME$>$ \\
    $<$LNAME$>$Ellis$<$/LNAME$>$ \\
  $<$/AU$>$ \\
  $<$AU AF=``2"$>$ \\
    $<$FNAME$>$Jean-Paul$<$/FNAME$>$ \\
    $<$LNAME$>$Kneib$<$/LNAME$>$ \\
  $<$/AU$>$ \\
  $<$AU AF=``2"$>$ \\
    $<$FNAME$>$Jean-Francois$<$/FNAME$>$ \\
    $<$LNAME$>$LeBorgne$<$/LNAME$>$ \\
  $<$/AU$>$ \\
  $<$AU AF=``2"$>$ \\
    $<$FNAME$>$Roser$<$/FNAME$>$ \\
    $<$LNAME$>$Pello$<$/LNAME$>$ \\
  $<$/AU$>$ \\
  $<$AU AF=``3"$>$ \\
    $<$FNAME$>$Ian$<$/FNAME$>$ \\
    $<$LNAME$>$Smail$<$/LNAME$>$ \\
  $<$/AU$>$ \\
  $<$AU AF=``4"$>$ \\
    $<$FNAME$>$Blai$<$/FNAME$>$ \\
    $<$LNAME$>$Sanahuja$<$/LNAME$>$ \\
  $<$/AU$>$ \\
$<$/AUTHORS$>$ \\
$<$AFFILIATIONS$>$ \\
  $<$AF ident=``AF\_1"$>$Cambridge, Univ.$<$/AF$>$ \\
  $<$AF ident=``AF\_2"$>$Observatoire Midi-Pyrenees$<$/AF$>$ \\
  $<$AF ident=``AF\_3"$>$Durham, Univ.$<$/AF$>$ \\
  $<$AF ident=``AF\_4"$>$Barcelona, Univ.$<$/AF$>$ \\
$<$/AFFILIATIONS$>$ \\
$<$MSTRING$>$Royal Astronomical Society, Monthly Notices, vol. 295, p. 75$<$/MSTRING$>$ \\
$<$MONOGRAPH$>$ \\
  $<$MTITLE$>$Royal Astronomical Society, Monthly Notices$<$/MTITLE$>$ \\
  $<$VOLUME$>$295$<$/VOLUME$>$ \\
$<$/MONOGRAPH$>$ \\
$<$PAGE$>$75$<$/PAGE$>$ \\
$<$PUBDATE$>$ \\
  $<$YEAR$>$1998$<$/YEAR$>$ \\
  $<$MONTH$>$03$<$/MONTH$>$ \\
$<$/PUBDATE$>$ \\
{\bf $<$CATEGORIES$>$ \\
  $<$CA$>$Astrophysics$<$/CA$>$ \\
$<$CATEGORIES$>$ \\
$<$BIBCODE$>$1998MNRAS.295...75E$<$/BIBCODE$>$ \\
$<$BIBTYPE$>$article$<$/BIBTYPE$>$ \\
$<$IDENTIFIERS$>$ \\
  $<$ID type=``ACCNO"$>$A98-51106$<$/ID$>$ \\
$<$/IDENTIFIERS$>$} \\
{\bf $<$KEYWORDS system=``STI"$>$ \\
$<$KW$>$GRAVITATIONAL LENSES$<$/KW$>$ \\
$<$KW$>$RED SHIFT$<$/KW$>$ \\
$<$KW$>$HUBBLE SPACE TELESCOPE$<$/KW$>$ \\
$<$KW$>$GALACTIC CLUSTERS$<$/KW$>$ \\
$<$KW$>$ASTRONOMICAL SPECTROSCOPY$<$/KW$>$ \\
$<$KW$>$MASS DISTRIBUTION$<$/KW$>$ \\
$<$KW$>$SPECTROGRAPHS$<$/KW$>$ \\
$<$KW$>$PREDICTION ANALYSIS TECHNIQUES$<$/KW$>$ \\
$<$KW$>$ASTRONOMICAL PHOTOMETRY$<$/KW$>$ \\
$<$/KEYWORDS$>$} \\
$<$ABSTRACT$>$ \\
We present deep spectroscopic measurements of 18 distant field galaxies
identified as gravitationally lensed arcs in a Hubble Space Telescope
image of the cluster Abell 2218. Redshifts of these objects were
predicted by Kneib et al. using a lensing analysis constrained by the
properties of two bright arcs of known redshift and other multiply
imaged sources. The new spectroscopic identifications were obtained
using long exposures with the LDSS-2 spectrograph on the William
Herschel Telescope, and demonstrate the capability of that instrument to
reach new limits, R = 24; the lensing magnification implies true source
magnitudes as faint as R = 25. Statistically, our measured redshifts are
in excellent agreement with those predicted from Kneib et al.'s lensing
analysis, and this gives considerable support to the redshift
distribution derived by the lensing inversion method for the more
numerous and fainter arclets extending to R = 25.5. We explore the
remaining uncertainties arising from both the mass distribution in the
central regions of Abell 2218 and the inversion method itself, and
conclude that the mean redshift of the faint field population at R =
25.5 (B = 26-27) is low, (z = 0.8-1). We discuss this result in the
context of redshift distributions estimated from multicolor photometry.\\
$<$\/ABSTRACT$>$ \\
$<$/BIBRECORD$>$ \\

\noindent
$<$BIBRECORD origin=``MNRAS"$>$ \\
$<$TITLE$>$Spectroscopic confirmation of redshifts 
predicted by gravitational lensing$<$/TITLE$>$ \\
{\bf $<$AUTHORS$>$ \\
  $<$AU AF=``1"$>$ \\
    $<$FNAME$>$Tim$<$/FNAME$>$ \\
    $<$LNAME$>$Ebbels$<$/LNAME$>$ \\
  $<$/AU$>$ \\
  $<$AU AF=``1" EM=``1"$>$ \\
    $<$FNAME$>$Richard$<$/FNAME$>$ \\
    $<$LNAME$>$Ellis$<$/LNAME$>$ \\
  $<$/AU$>$ \\
  $<$AU AF=``2"$>$ \\
    $<$FNAME$>$Jean-Paul$<$/FNAME$>$ \\
    $<$LNAME$>$Kneib$<$/LNAME$>$ \\
  $<$/AU$>$ \\
  $<$AU AF=``2"$>$ \\
    $<$FNAME$>$Jean-Fran\&ccedil;ois$<$/FNAME$>$ \\
    $<$LNAME$>$LeBorgne$<$/LNAME$>$ \\
  $<$/AU$>$ \\
  $<$AU AF=``2"$>$ \\
    $<$FNAME$>$Roser$<$/FNAME$>$ \\
    $<$LNAME$>$Pell\&oacute;$<$/LNAME$>$ \\
  $<$/AU$>$ \\
  $<$AU AF=``3"$>$ \\
    $<$FNAME$>$Ian$<$/FNAME$>$ \\
    $<$LNAME$>$Smail$<$/LNAME$>$ \\
  $<$/AU$>$ \\
  $<$AU AF=``4"$>$ \\
    $<$FNAME$>$Blai$<$/FNAME$>$ \\
    $<$LNAME$>$Sanahuja$<$/LNAME$>$ \\
  $<$/AU$>$ \\
$<$/AUTHORS$>$ \\
$<$AFFILIATIONS$>$ \\
  $<$AF ident=``AF\_1"$>$Institute of Astronomy, Madingley Road, Cambridge CB3 0HA$<$/AF$>$ \\
  $<$AF ident=``AF\_2"$>$Observatoire Midi-Pyr\&eacute;n\&eacute;es, 14 Avenue E. Belin$<$/AF$>$ \\
  $<$AF ident=``AF\_3"$>$Department of Physics, University of Durham, South Road, Durham DH1 3LE$<$/AF$>$ \\
  $<$AF ident=``AF\_4"$>$Departament d\&apos;Astronomia i Meteorologia, Universitat de Barcelona, Diagonal 648, 08028 Barcelona, Spain$<$/AF$>$ \\
$<$/AFFILIATIONS$>$ \\
$<$EMAILS$>$ \\
  $<$EM ident=``EM\_1"$>$rse$@$ast.cam.ac.uk$<$/EM$>$ \\
$<$/EMAILS$>$ \\
$<$MSTRING$>$Monthly Notices of the Royal Astronomical Society, Volume 295, Issue 1, pp. 75-91.$<$/MSTRING$>$ \\
$<$MONOGRAPH$>$ \\
  $<$MTITLE$>$Monthly Notices of the Royal Astronomical Society$<$/MTITLE$>$ \\
  $<$MTITLE$>$Monthly Notices of the Royal Astronomical Society$<$/MTITLE$>$ \\
  $<$VOLUME$>$295$<$/VOLUME$>$ \\
  $<$ISSUE$>$1$<$/ISSUE$>$ \\
$<$/MONOGRAPH$>$ \\
$<$PAGE$>$75$<$/PAGE$>$ \\
$<$LPAGE$>$91$<$/LPAGE$>$ \\
$<$PUBDATE$>$ \\
  $<$YEAR$>$1998$<$/YEAR$>$ \\
  $<$MONTH$>$03$<$/MONTH$>$ \\
$<$/PUBDATE$>$ \\
$<$COPYRIGHT$>$1998: The Royal Astronomical Society$<$/COPYRIGHT$>$} \\ 
$<$BIBCODE$>$1998MNRAS.295...75E$<$/BIBCODE$>$ \\
$<$KEYWORDS system=``AAS"$>$ \\
  $<$KW$>$GALAXIES: CLUSTERS: INDIVIDUAL: ABELL 2218$<$/KW$>$ \\
  $<$KW$>$GALAXIES: EVOLUTION$<$/KW$>$ \\
  $<$KW$>$COSMOLOGY: OBSERVATIONS$<$/KW$>$ \\
  $<$KW$>$GRAVITATIONAL LENSING$<$/KW$>$ \\
$<$/KEYWORDS$>$ \\
{\bf $<$ABSTRACT$>$ \\
We present deep spectroscopic measurements of 18 distant field galaxies
identified as gravitationally lensed arcs in a Hubble Space Telescope
image of the cluster Abell2218. Redshifts of these objects were
predicted by Kneib et al. using a lensing analysis constrained by the
properties of two bright arcs of known redshift and other multiply
imaged sources. The new spectroscopic identifications were obtained
using long exposures with the LDSS-2 spectrograph on the William
Herschel Telescope, and demonstrate the capability of that instrument to
reach new limits, R\&sime;24 the lensing magnification implies true source
magnitudes as faint as R\&sime;25. Statistically, our measured redshifts are
in excellent agreement with those predicted from Kneib et al.'s lensing
analysis, and this gives considerable support to the redshift
distribution derived by the lensing inversion method for the more
numerous and fainter arclets extending to R\&sime;25.5. We explore the
remaining uncertainties arising from both the mass distribution in the
central regions of Abell2218 and the inversion method itself, and
conclude that the mean redshift of the faint field population at R\&sime;25.5
(B\&sim;26\&ndash;27) is low, \&lang;z\&rang;=0.8\&ndash;1. We discuss this result 
in the context of redshift distributions estimated from multicolour photometry. 
Although such comparisons are not straightforward, we suggest that photometric
techniques may achieve a reasonable level of agreement, particularly
when they include near-infrared photometry with discriminatory
capabilities in the 1\&lt;z\&lt;2 range. \\
$<$/ABSTRACT$>$} \\
$<$/BIBRECORD$>$ \\

\noindent
$<$BIBRECORD origin=``SIMBAD"$>$ \\
$<$TITLE$>$Spectroscopic confirmation of redshifts 
predicted by gravitational lensing.$<$/TITLE$>$ \\
$<$AUTHORS$>$ \\
  $<$AU$>$ \\
    $<$FNAME$>$T.$<$/FNAME$>$ \\
    $<$LNAME$>$Ebbels$<$/LNAME$>$ \\
  $<$/AU$>$ \\
  $<$AU$>$ \\
    $<$FNAME$>$R.$<$/FNAME$>$ \\
    $<$LNAME$>$Ellis$<$/LNAME$>$ \\
  $<$/AU$>$ \\
  $<$AU$>$ \\
    $<$FNAME$>$J.-P.$<$/FNAME$>$ \\
    $<$LNAME$>$Kneib$<$/LNAME$>$ \\
  $<$/AU$>$ \\
  $<$AU$>$ \\
    $<$FNAME$>$J.-F.$<$/FNAME$>$ \\
    $<$LNAME$>$LeBorgne$<$/LNAME$>$ \\
  $<$/AU$>$ \\
  $<$AU$>$ \\
    $<$FNAME$>$R.$<$/FNAME$>$ \\
    $<$LNAME$>$Pell\&oacute;$<$/LNAME$>$ \\
  $<$/AU$>$ \\
  $<$AU$>$ \\
    $<$FNAME$>$I.$<$/FNAME$>$ \\
    $<$LNAME$>$Smail$<$/LNAME$>$ \\
  $<$/AU$>$ \\
  $<$AU$>$ \\
    $<$FNAME$>$B.$<$/FNAME$>$ \\
    $<$LNAME$>$Sanahuja$<$/LNAME$>$ \\
  $<$/AU$>$ \\
$<$/AUTHORS$>$ \\
$<$MSTRING$>$Mon. Not. R. Astron. Soc., 295, 75-91 (1998)$<$/MSTRING$>$ \\
$<$MONOGRAPH$>$ \\
  $<$MTITLE$>$Mon. Not. R. Astron. Soc.$<$/MTITLE$>$ \\
  $<$VOLUME$>$295$<$/VOLUME$>$ \\
$<$MONOGRAPH$>$ \\
$<$PAGE$>$75$<$/PAGE$>$ \\
$<$LPAGE$>$91$<$/LPAGE$>$ \\
$<$PUBDATE$>$ \\
  $<$YEAR$>$1998$<$/YEAR$>$ \\
  $<$MONTH$>$03$<$/MONTH$>$ \\
$<$/PUBDATE$>$ \\
$<$BIBCODE$>$1998MNRAS.295...75E$<$/BIBCODE$>$ \\
$<$/BIBRECORD$>$ \\
$<$/ADS\_BIBALL$>$ \\

\section {\label {appendixc}}

An example of an extracted text file from the ADS Abstract Service showing 
only the preferred instances of each field in XML markup for same bibliographic entry listed in Appendix B.

\noindent
$<$?xml version=``1.0"?$>$ \\
$<$!DOCTYPE ADS\_ABSTRACT SYSTEM ``ads.dtd"$>$ \\
$<$ADS\_ABSTRACT$>$ \\

\noindent
$<$TITLE$>$Spectroscopic confirmation of redshifts 
predicted by gravitational lensing$<$/TITLE$>$ \\
$<$AUTHORS$>$ \\
  $<$AU AF=``1"$>$ \\
    $<$FNAME$>$Tim$<$/FNAME$>$ \\
    $<$LNAME$>$Ebbels$<$/LNAME$>$ \\
  $<$/AU$>$ \\
  $<$AU AF=``1" EM=``1"$>$ \\
    $<$FNAME$>$Richard$<$/FNAME$>$ \\
    $<$LNAME$>$Ellis$<$/LNAME$>$ \\
  $<$/AU$>$ \\
  $<$AU AF=``2"$>$ \\
    $<$FNAME$>$Jean-Paul$<$/FNAME$>$ \\
    $<$LNAME$>$Kneib$<$/LNAME$>$ \\
  $<$/AU$>$ \\
  $<$AU AF=``2"$>$ \\
    $<$FNAME$>$Jean-Fran\&ccedil;ois$<$/FNAME$>$ \\
    $<$LNAME$>$LeBorgne$<$/LNAME$>$ \\
  $<$/AU$>$ \\
  $<$AU AF=``2"$>$ \\
    $<$FNAME$>$Roser$<$/FNAME$>$ \\
    $<$LNAME$>$Pell\&oacute;$<$/LNAME$>$ \\
  $<$/AU$>$ \\
  $<$AU AF=``3"$>$ \\
    $<$FNAME$>$Ian$<$/FNAME$>$ \\
    $<$LNAME$>$Smail$<$/LNAME$>$ \\
  $<$/AU$>$ \\
  $<$AU AF=``4"$>$ \\
    $<$FNAME$>$Blai$<$/FNAME$>$ \\
    $<$LNAME$>$Sanahuja$<$/LNAME$>$ \\
  $<$/AU$>$ \\
$<$/AUTHORS$>$ \\
$<$AFFILIATIONS$>$ \\
  $<$AF ident=``AF\_1"$>$Institute of Astronomy, Madingley Road, Cambridge CB3 0HA$<$/AF$>$ \\
  $<$AF ident=``AF\_2"$>$Observatoire Midi-Pyr\&eacute;n\&eacute;es, 14 Avenue E. Belin$<$/AF$>$ \\
  $<$AF ident=``AF\_3"$>$Department of Physics, University of Durham, South Road, Durham DH1 3LE$<$/AF$>$ \\
  $<$AF ident=``AF\_4"$>$Departament d\&apos;Astronomia i Meteorologia, Universitat de Barcelona, Diagonal 648, 08028 Barcelona, Spain$<$/AF$>$ \\
$<$/AFFILIATIONS$>$ \\
$<$EMAILS$>$ \\
  $<$EM ident=``EM\_1"$>$rse$@$ast.cam.ac.uk$<$/EM$>$ \\
$<$/EMAILS$>$ \\
$<$MSTRING$>$Monthly Notices of the Royal Astronomical Society, Volume 295, Issue 1, pp. 75-91.$<$/MSTRING$>$ \\
$<$MONOGRAPH$>$ \\
  $<$MTITLE$>$Monthly Notices of the Royal Astronomical Society$<$/MTITLE$>$ \\
  $<$VOLUME$>$295$<$/VOLUME$>$ \\
  $<$ISSUE$>$1$<$/ISSUE$>$ \\
$<$/MONOGRAPH$>$ \\
$<$PAGE$>$75$<$/PAGE$>$ \\
$<$LPAGE$>$91$<$/LPAGE$>$ \\
$<$PUBDATE$>$ \\
  $<$YEAR$>$1998$<$/YEAR$>$ \\
  $<$MONTH$>$03$<$/MONTH$>$ \\
$<$/PUBDATE$>$ \\
$<$CATEGORIES$>$ \\
  $<$CA$>$Astrophysics$<$/CA$>$ \\
$<$CATEGORIES$>$ \\
$<$COPYRIGHT$>$1998: The Royal Astronomical Society$<$/COPYRIGHT$>$ \\ 
$<$IDENTIFIERS$>$ \\
  $<$ID type=``ACCNO"$>$A98-51106$<$/ID$>$ \\
$<$/IDENTIFIERS$>$ \\
$<$ORIGINS$>$ \\
  $<$OR$>$STI$<$/OR$>$ \\
  $<$OR$>$MNRAS$<$/OR$>$ \\
  $<$OR$>$SIMBAD$<$/OR$>$ \\
$<$/ORIGINS$>$ \\
$<$BIBCODE$>$1998MNRAS.295...75E$<$/BIBCODE$>$ \\
$<$BIBTYPE$>$article$<$/BIBTYPE$>$ \\
$<$KEYWORDS SYSTEM=``STI"$>$ \\
  $<$KW$>$GRAVITATIONAL LENSES$<$/KW$>$ \\
  $<$KW$>$RED SHIFT$<$/KW$>$ \\
  $<$KW$>$HUBBLE SPACE TELESCOPE$<$/KW$>$ \\
  $<$KW$>$GALACTIC CLUSTERS$<$/KW$>$ \\
  $<$KW$>$ASTRONOMICAL SPECTROSCOPY$<$/KW$>$ \\
  $<$KW$>$MASS DISTRIBUTION$<$/KW$>$ \\
  $<$KW$>$SPECTROGRAPHS$<$/KW$>$ \\
  $<$KW$>$PREDICTION ANALYSIS TECHNIQUES$<$/KW$>$ \\
  $<$KW$>$ASTRONOMICAL PHOTOMETRY$<$/KW$>$ \\
$<$/KEYWORDS$>$ \\
$<$KEYWORDS SYSTEM=``AAS"$>$ \\
  $<$KW$>$GALAXIES: CLUSTERS: INDIVIDUAL: ABELL 2218$<$/KW$>$ \\
  $<$KW$>$GALAXIES: EVOLUTION$<$/KW$>$ \\
  $<$KW$>$COSMOLOGY: OBSERVATIONS$<$/KW$>$ \\
  $<$KW$>$GRAVITATIONAL LENSING$<$/KW$>$ \\
$<$/KEYWORDS$>$ \\
$<$ABSTRACT$>$ \\
We present deep spectroscopic measurements of 18 distant field galaxies
identified as gravitationally lensed arcs in a Hubble Space Telescope
image of the cluster Abell2218. Redshifts of these objects were
predicted by Kneib et al. using a lensing analysis constrained by the
properties of two bright arcs of known redshift and other multiply
imaged sources. The new spectroscopic identifications were obtained
using long exposures with the LDSS-2 spectrograph on the William
Herschel Telescope, and demonstrate the capability of that instrument to
reach new limits, R\&sime;24 the lensing magnification implies true source
magnitudes as faint as R\&sime;25. Statistically, our measured redshifts are
in excellent agreement with those predicted from Kneib et al.'s lensing
analysis, and this gives considerable support to the redshift
distribution derived by the lensing inversion method for the more
numerous and fainter arclets extending to R\&sime;25.5. We explore the
remaining uncertainties arising from both the mass distribution in the
central regions of Abell2218 and the inversion method itself, and
conclude that the mean redshift of the faint field population at R\&sime;25.5
(B\&sim;26\&ndash;27) is low, \&lang;z\&rang;=0.8\&ndash;1. We discuss this result 
in the context of redshift distributions estimated from multicolour photometry. 
Although such comparisons are not straightforward, we suggest that photometric
techniques may achieve a reasonable level of agreement, particularly
when they include near-infrared photometry with discriminatory
capabilities in the 1\&lt;z\&lt;2 range. \\
$<$/ABSTRACT$>$ \\
$<$/ADS\_ABSTRACT$>$ \\

\end{document}